\documentclass{iau} 
\usepackage{graphicx}

\title[IAUS 346]
{Constraining the progenitor evolution of GW 150914}

\author[Jorick S,. Vink]   
{Jorick S. Vink$^1$}

\affiliation{$^1$Armagh Observatory and Planetarium, BT61 9DG Armagh, College Hill, Northern Ireland \\email: {\tt jorick.vink@armagh.ac.uk}}

\pubyear{2018}
\volume{346}  
\jname{Title of your IAU Symposium}
\editors{A.C. Editor, B.D. Editor \& C.E. Editor, eds.}
\begin{document}

\maketitle

\begin{abstract}
One of the largest surprises from the LIGO results regarding the first gravitational wave detection (GW 150914) 
was the fact the black holes (BHs) were "heavy", of order 30 - 40 $M_{\odot}$. 
The most promising explanation for this obesity is that the BH-BH merger occurred at low metallicity ($Z$): when the iron (Fe) contents is lower this is expected to result 
in weaker mass loss during the Wolf-Rayet (WR) phase. We therefore critically evaluate the claims for the reasons of heavy BHs as a function of $Z$ in the literature.
Furthermore, weaker stellar winds might lead to more rapid stellar rotation, allowing WR and BH progenitor evolution in a chemically homogeneous manner. 
However, there is as yet no empirical evidence for more rapid rotation amongst WR stars in the low $Z$ environment of the Magellanic Clouds. 
Due to the intrinsic challenge of determining WR rotation rates from emission lines, the most promising avenue to constrain rotation-rate distributions amongst various WR subgroups 
is through the utilisation of their emission lines in polarised light. We thus provide an overview of linear spectro-polarimetry observations of both single and binary WRs in the Galaxy, as well as 
the Large and Small Magellanic Clouds, at 50\%  and 20\% of solar $Z$, respectively. Initial results suggest that the route of chemically homogeneous evolution (CHE) through stellar 
rotation is challenging, whilst the alternative of a post-LBV or common envelope evolution is more likely.
\keywords{gravitational waves, polarization, stars: early-type, stars: evolution, stars: mass loss, stars: rotation, stars: winds, outflows, stars: Wolf-Rayet}
\end{abstract}

\firstsection 
\section{Introduction}

One of the main surprises regarding the first gravitational wave detections by 
LIGO concerning the physical merging of 2 black holes (BHs) was the fact that the masses 
inferred for these BHs were very heavy -- of order 30 - 40 $M_{\odot}$ (Abbott et al. 2016). 
Within our own Galactic environment the maximum black hole mass is thought to be of 
order 10 $M_{\odot}$ (Belczynski et al. 2010).  

For this reason there are 2 possible solutions to the problem. Either the GW150914 event took place
in an environment that was low metallicity, reducing the amount of mass loss during stellar evolution,
or the initial stars started off with very high masses. 
Even if there had been relatively little mass loss, the initial 
masses must have been higher than 40 $M_{\odot}$, relating to a stellar mass regime where winds are 
a critical ingredient for massive-star evolution (e.g. Higgins \& Vink 2018).  For very massive stars (VMS), defined 
with masses over $\simeq$100 $M_{\odot}$ (Vink et al. 2015), stellar winds completely dominate 
their evolution and fate (Woosley \& Heger 2015; Hirschi 2015). 

In this contribution, we constrain the physics and evolution of the progenitor
of GW150914 in terms of the geometry and the amount of the mass loss from Wolf-Rayet (WR) stars, which 
are thought to be the direct progenitors of BHs.

\begin{figure}[b]
\begin{center}
 \includegraphics[width=4.8in]{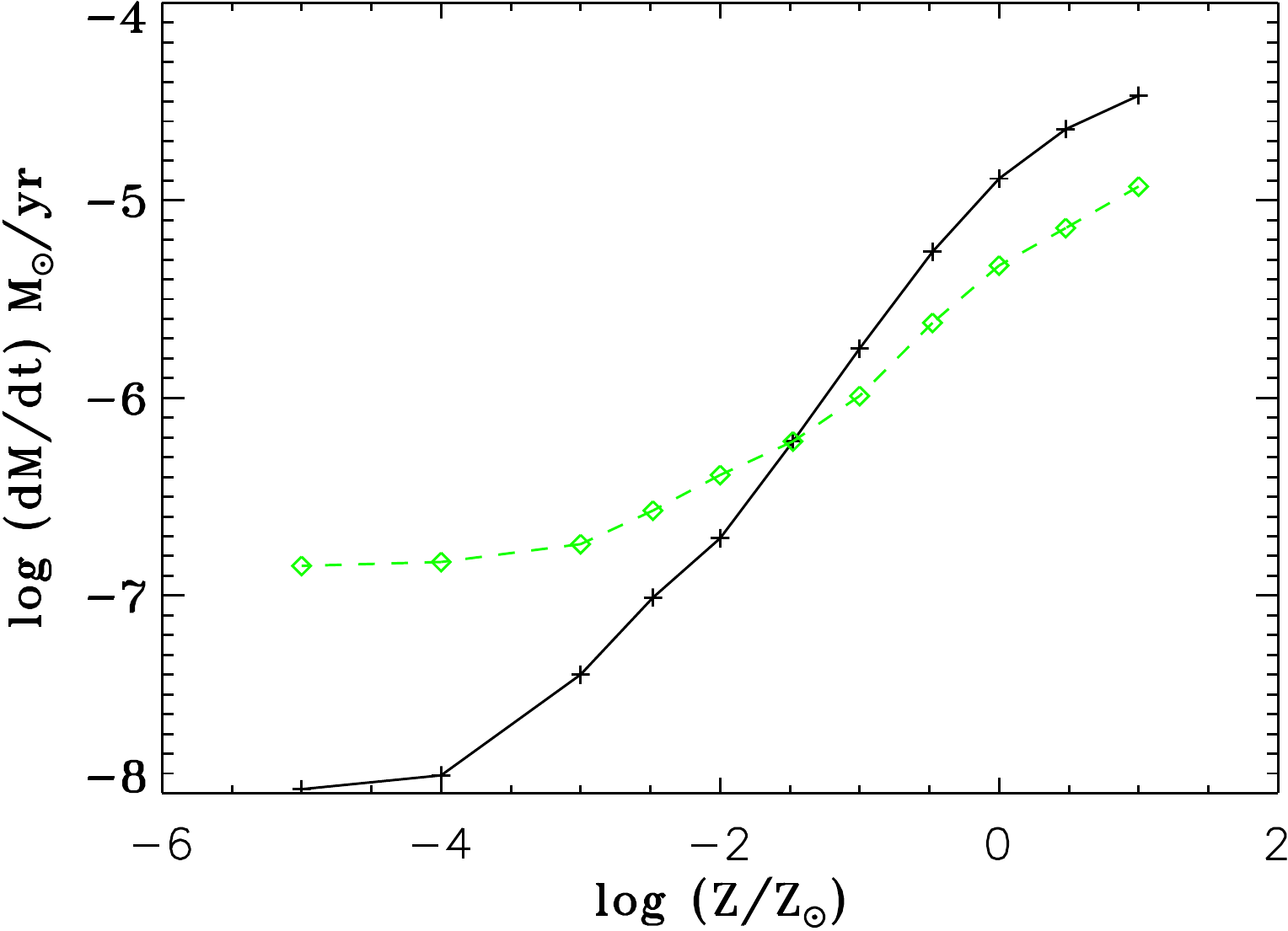} 
  \caption{Predicted mass-loss rates of WR stars versus host galaxy metallicity. The black solid line indicates the steeper dependence 
 for the nitrogen-rioch WN stars, whilst the green dashed line indicates the shallower slope for the carbon-rich WC stars. From Vink
 \& de Koter (2005).}
   \label{fig1}
\end{center}
\end{figure}

\section{Stellar winds at low metallicity}

Whilst it has been known for decades that stellar winds during the initial O star phase depend 
on metallicity $Z$ (Abbott et al. 1982; Kudritzki et al. 1987; Vink et al. 2001), the realisation that
WR stars also depend on the iron (Fe) contents of the host galaxy is rather more recent (Vink \& de Koter 2005; 
Hainich et al. 2015). Until 2005, most stellar evolution modellers assumed that the host metallicity (Fe) was 
so low in terms of stellar abundance when compared to the self-enriched carbon during the WC
phase that mass loss was assumed to be independent of host galaxy metallicity.

\begin{figure}[b]
\begin{center}
 \includegraphics[width=5.4in]{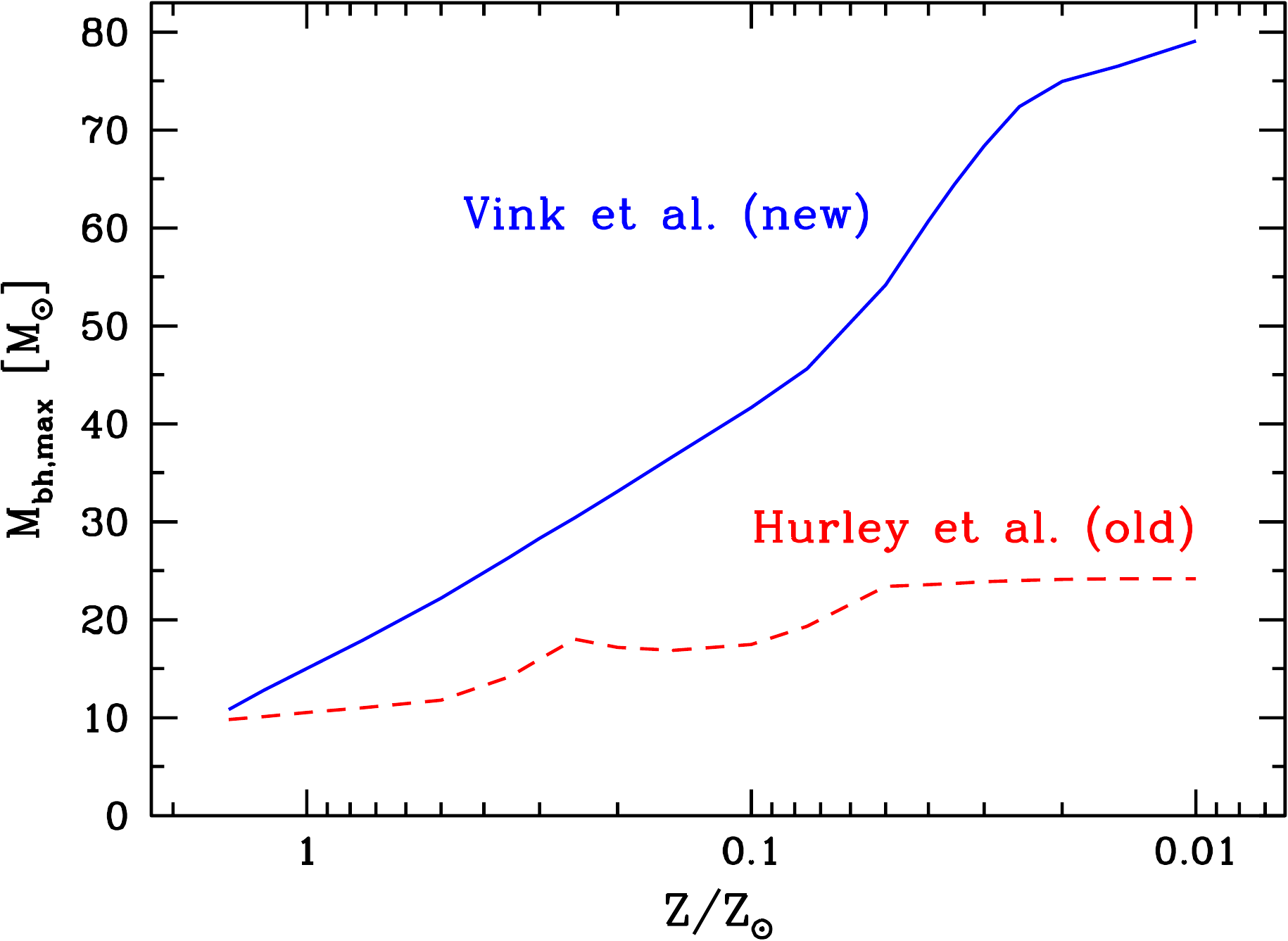} 
 \caption{Predicted maximum BH mass versus metallicity. The red dashed line shows the almost $Z$ independent maximum BH mass
 when including the situation around the year 2000 (Hurley et al. 2000), whilst including the $Z$ dependence of not only
 the Vink et al. (2001) O-type stars, but also the WR $Z$ dependent winds of Vink \& de Koter (2005) leads to a strong
 sensitivity of maximum BH mass on $Z$, as indicated by the blue solid line. From Belczynski et al. (2010). The same plot
 with incorrect labels was included in Abbott et al. (2016).}
   \label{fig2}
\end{center}
\end{figure}

Using an established Monte Carlo method, Vink \& de Koter (2005) showed that WR mass-loss rates depend on the 
host galaxy metallicity (Fe) after all.  Figure 1 shows the predicted WR mass-loss versus $Z$ dependence for both
nitrogen-rich WN stars (in black solid) and the slightly shallower $Z$ dependence for 
carbon-rich WC stars (in dashed green). Note that the self-enriched materials are not taken into account in the definition 
of $Z$ on the x-axis. In more recent times, Vink (2017) confirmed the theoretical $Z_{\rm Fe}$ dependence for optically thin stripped 
helium stars using a dynamically consistent version of the Monte Carlo method (M\"uller \& Vink 2008).

A direct consequence of including $Z$ dependent mass-loss rates of WR stars into massive star evolution models 
was the prediction of more massive BHs (Eldridge \& Vink 2006). Figure 2 shows the expected maximum BH mass 
at a given host galaxy metallicity from a single star population synthesis by Belczynski et al. (2010). The dashed red
line indicates the older view that independent of host galaxy $Z$ the maximum BH mass would not exceed $\sim$10-20 $M_{\odot}$,
as the WR mass-loss rates were thought to be due to self-enrichment by carbon and oxygen. Only when including the 
newer Vink \& de Koter (2005) WR Fe dependent mass-loss rates one may expect to find heavier BHs at lower metallicities. 

Exactly the same plot was also included in the astrophysical interpretation
paper of GW 150914 by the LIGO consortium (Abbott et al. 2016) but, confusingly, the older and newer expectations 
of WR mass loss versus $Z$ implementations were here referred to as 'stronger' and 'weaker' stellar winds. That is not correct, as 
absolute values of the mass-loss rates have not changed. Instead, it is the weaker mass loss at lower $Z$ that is the key physics 
explaining the heavier BHs at low $Z$. At the same time, there have not been any substantial changes at solar $Z$, as can
be seen at the intersection of the 2 curves on the left-hand side of Figure 2.

Simply reading off the y-axis of Figure 2 for  a 40 $M_{\odot}$ maximum BH mass, immediately leads to the conclusion that the 
chemical environment of the GW150914 progenitor should have been 1/10 $Z_{\odot}$ or less. Detailed single star or binary evolution 
does not appear to be particularly relevant for making this particular inference. 
Instead, given the intrinsically low $Z$, GW150914-type events can teach us interesting physics about winds and stellar 
evolution at low $Z$.

\section{Interplay between rotation and winds}

In order to not only predict compact object masses, but also their spin, it is important to consider the theory of stellar rotation 
and winds. The first aspect is that the relation can work in both directions: stellar winds may remove angular momentum, thus 
braking the star, even in the absence of a magnetic field (e.g. Langer 2012), but reversely it has been argued that stellar rotation
may increase the overall mass-loss rate (e.g. Maeder \& Meynet 2000).  In more recent times, 2D dynamical calculations 
by M\"uller \& Vink (2014) showed there are cases where the overall mass-loss rate may actually decrease with respect to their 
non-rotating counterparts.

The third relevant aspect is to consider the geometry of 
rotating winds: do we expect equatorial enhancement (which may remove angular momentum very efficiently) or polar winds?
Let us briefly review the key physical ingredients. 
Friend \& Abbott (1986) argued that as a result of a lower effective gravity from a rotating star, mass loss from the equator would 
be more efficient than from the pole. Later, Cranmer \& Owocki (1995) found that due to the Von Zeipel gravity darkening the pole would be 
brighter and thus mass would preferentially be lost from the pole instead. Alternatively, taking both these competing effects into account, for certain temperatures around the 
bi-stability jump temperature, mass would predominately be lost from the equator after all (Pelupessy et al. 2000). It is clear 
that the situation regarding the geometry is complex, and that theory needs guidance from observations in order to make 
progress. 

\section{Testing stellar rotation with polarimetry}

\begin{figure}[b]
\begin{center}
 \includegraphics[width=6.0in]{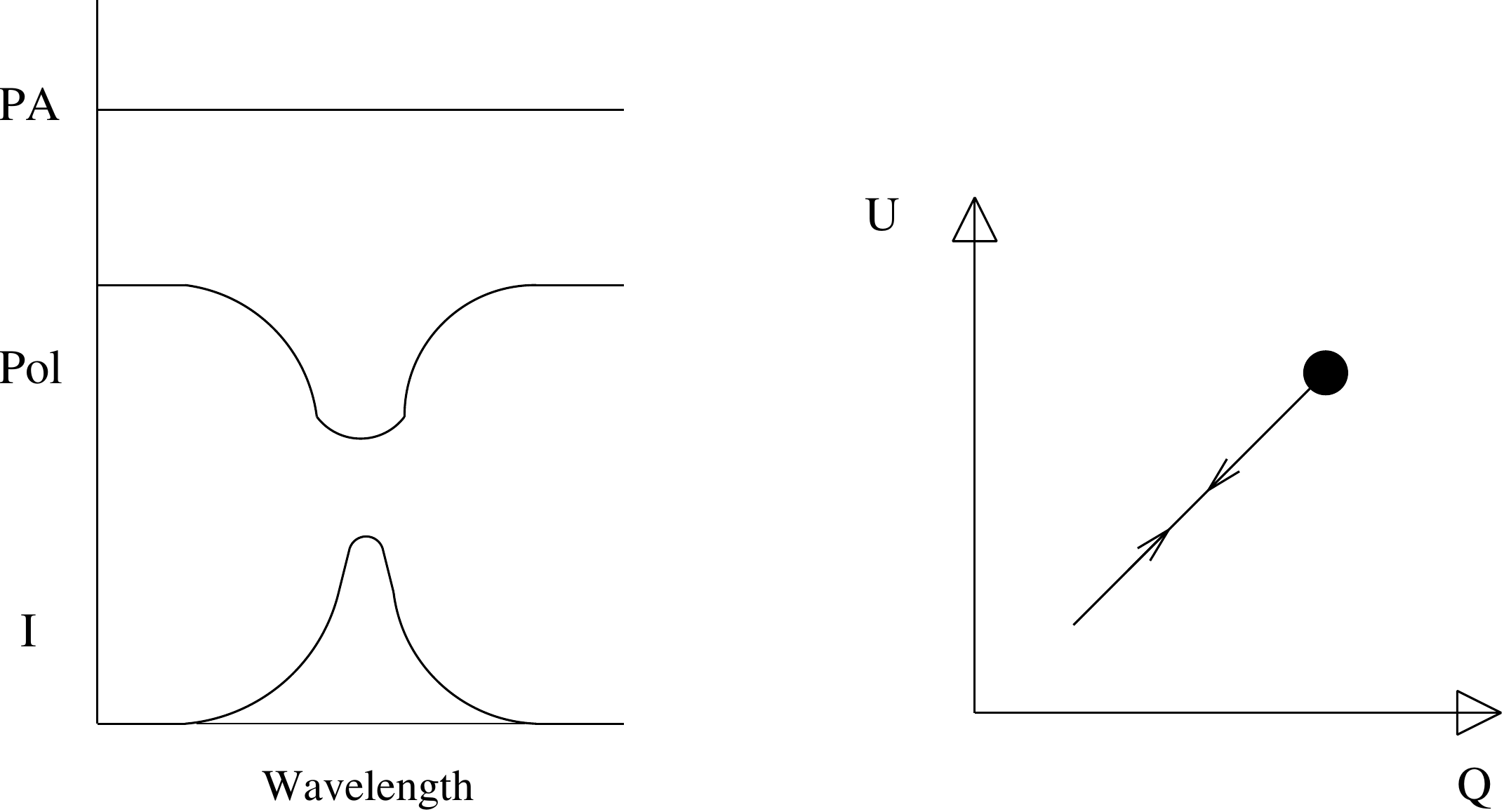} 
 \caption{Polarization triplot (on the left-hand side) and the 
Stokes QU plane (right-hand side). For an emission line (lower panel triplot) one expects a 'depolarization' in the middle 
panel of the triplot if the innermost geometry is flattened. For a simple disk ,the position angle (PA) of the
disk remains constant, and the PA can also be read off from the QU plane (right-hand side). }
   \label{fig3}
\end{center}
\end{figure}

Until astronomers are able to spatially resolve the innermost radii of hot massive stars, linear spectropolarimetry is the only
tool available to dissect geometry of stellar winds. The technique was already applied in the 1970s for classical Be stars, and 
was later also applied to young pre-main sequence Herbig Ae/Be and T Tauri stars to understand accretion disks. 
One of the key strengths of {\it spectro}polarimetry is that there is no dependence on any dust particles between the 
star under consideration and the observer (due to interstellar polarization).  

Figure 3 shows the expectations from a disk around a star in the polarization triplot (on the left-hand side) and the 
Stokes QU plane. For an emission line (lower panel triplot), one only expects to see a 'depolarization' in the middle 
panel of the triplot if the innermost geometry is significantly flattened, i.e. disk-like. For a simple disk, the position angle (PA) remains constant, and the PA can also be determined from the QU plane (see the right-hand side). 

Given the expected lower mass-loss rates of WR stars at lower $Z$, one might perhaps expect disks to be 
more prevalent at lower $Z$ than in our Milky Way. Indeed many B-type stars in the SMC seem to be Be stars (e.g. Castro et al. 2018).
However, in our recent VLT-FORS polarisation study of large samples of WR stars in the low $Z$ environments of the 
Magallanic Clouds (39 in the LMC; 
all 12 WRs in the SMC) we found the incidence of depolarisation 'line effects' (Fig. 3) to be indistinguishable from those in the 
Milky Way (Vink \& Harries 2017). 

This appears to be quite a revelation for stellar modellers attempting to explain the evolution of GW 150914 with physics related to rapid rotation, such as the rotationally-induced chemically homogeneous evolution (CHE), as 
we have basically no empirical evidence that WR stars at low metallicity rotate any faster than those in the Milky Way.

\section{Remarks on evolutionary scenarios}

The results of the VLT study of Vink \& Harries (2017) suggest that WR stars in low $Z$ environments do not 
rotate any faster than at high $Z$. What does this imply for the subsequent evolution of WR stars into BHs, and how 
can this information be used to constrain the evolution towards the WR phase?

Independent of any subtle binary evolution effects that will undoubtedly come into play when we wish to explain the 
compact object mass spectrum, we can already learn some lessons regarding the physics of either component in the 
merging BH binary progenitor. 

First of all, the concept of CHE for the formation of WR stars at low $Z$ due to rotationally-induced CHE (Yoon \& Langer 
2005) is challenging to entertain given the lack of evidence for WR rotation at low $Z$. 
This would therefore also make it harder for this evolutionary pathway to be relevant for 
the very specific event of GW150914 (e.g. 
Mandel \& de Mink 2016; Marchant et al. 2016), unless such events 
are evolutionary unrelated to ordinary single star and binary evolution at high and low $Z$, as our VLT sample contained 
both single  WRs and binaries, and no difference in the line-effect frequency was noted.

Vink et al. (2011) argued that the most likely explanation for the fact that only a 10-20\% sub-population of WR stars 
is found to rotate is related to a more classical post-LBV like evolutionary channel. 
In the context of binary evolution 
we might translate this to common-envelope evolution (Belczynski et al. 2016), although the interesting discussion on CHE 
should certainly continue!

\end{document}